# Investigation of the Paschen Curve for Various Electrode Geometries in IEC Fusion Devices through Monte Carlo Simulations

William Bowers[1*], Sophie Gershaft[1,2]

[1] Eastside Preparatory School, Kirkland, WA, USA
[2] California Institute of Technology, Pasadena, CA, USA
[*] Corresponding author (whbowers06@gmail.com)

**Abstract**

The creation of plasma is key for achieving fusion in Inertial Electrostatic Confinement (IEC) fusion devices, and the conditions for such electrical breakdown are modelled by Paschen's law, which predicts the breakdown voltage of a system as a function of the product of pressure and gap distance (pd) between electrodes. However, the Paschen curve only models parallel plate configurations of electrodes and is rarely explored in the more complex electrode geometries often seen in IEC fusion devices, including Farnsworth-Hirsch fusors. To bridge this gap, we study the Paschen curve for various electrode configurations by use of Monte Carlo (MC) simulations in Mathematica and Java. We compute alpha–the rate of ionizations per unit length–in constant E-field systems to explore differences between predicted alpha values from the scientific literature and those from our MC simulations. We explore a Markov chain model for 1D alphas which outperforms the literature alpha prediction in modelling our 1D MC-derived alphas and more closely matches parallel plate MC breakdown results at the minimum. 3D simulations and breakdown plots are created for concentric sphere and sphere-in-cylinder geometries. We observe a widened breakdown curve for concentric spheres and sphere-in-cylinder geometries and note the reduced growth of breakdown voltage at small pd values for the sphere-in-cylinder case. These findings help explain the data collected from our own experimental fusor setup and suggest further work in the simulation of more complex electrode configurations and the incorporation of experimental data to verify results.

Keywords: Fusion, Plasma, Paschen's Law

## Introduction

The Paschen curve (figure 1, eq. 1) describes the conditions at which a gas undergoes electrical breakdown. Discovered by Friedrich Paschen in 1889, the law describes breakdown voltage, $V_B$, as a function of the product of pressure and gap distance between two electrodes (pd).

$$V = \frac{Apd}{\ln(pd) + B} \quad (1)$$

Here, $A = \frac{V_{min}}{(pd)_{min}}$, and $B = 1 - \ln((pd)_{min})$, where $V_{min}$ is the minimum breakdown voltage of the system and $(pd)_{min}$ is the minimum $pd$ where it occurs [1-2]. Incorporated into the $B$ parameter is another parameter, $\gamma_e$, which represents the minimum number of ionizations needed per electron for discharge to occur.

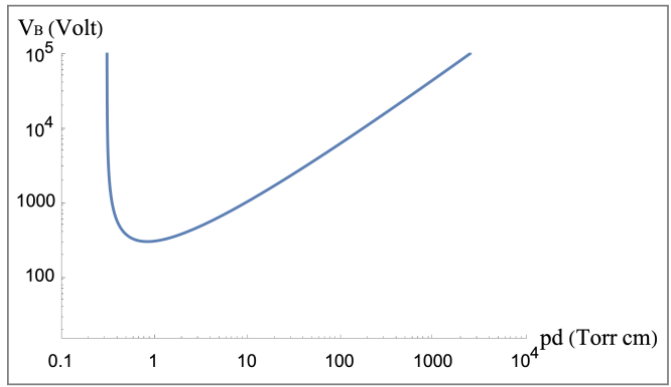

**Figure 1.** Paschen curve in air [3].

Paschen's model for breakdown describes the creation of plasma in a vacuum chamber consisting of a parallel configuration of two planar electrodes, which represents a homogeneous electric field. In the derivation of Paschen's law, the rate of ionizations per unit length, $\alpha$, is directly related to the electric field and is assumed to be constant. It is previously established that $\alpha = \frac{1}{\lambda} e^{-\frac{\lambda_i}{\lambda}}$, where $\lambda$ is the mean free path (MFP) and $\lambda_i$ is the ionizing distance, or the minimum distance that an electron must travel before gaining ionizing energy. This alpha assumption fails to describe electrode configurations with non-homogenous electric fields, where $\alpha$ varies with position. In addition, the literature derivation assumes that the current density at the anode is equal to the current density at the cathode, which is inaccurate for concentric sphere and sphere-in-cylinder geometries where the current density changes in proportion to the increase in the surface area from the cathode to the anode [1].

Multiple related but independent studies found that Paschen curves obtained by calculations and simulations in various setups were not in good agreement with experimental results but were well approximated by a modified Paschen's law [4-6]. One such study by McAllister [6] investigated Paschen curves for a strongly electronegative gas and found that breakdown voltage values fit a Paschen-like curve which does not fit Paschen's law, as shown by a linear regression analysis of the data. Lyu [4] demonstrated that experimental data for short glow discharge in grid and solid parallel plate electrodes was well approximated by a modified Paschen curve. Carey [7] also confirmed with experimental data that the breakdown voltage is not well modelled by the theoretical Paschen curve at high pd values. While multiple studies investigated the conditions under which Paschen's law deviates from experimental and simulated data in planar electrodes, more work was needed to study the breakdown voltage in concentric and coaxial electrodes of various geometries.

We research Paschen's law to describe the results observed in our experimental fusion reactor setup. In 2015, high school students at Eastside Preparatory School in Kirkland, WA, began working on a project of creating a Farnsworth-Hirsch nuclear fusion reactor in the school's technology laboratory. The experimental setup consists of a near-spherical hollow cathode held at a negative DC voltage, surrounded by a cylindrical anode maintained at ground (figure 2). The anode is a steel vacuum chamber with an inner diameter of 14.6 cm and an inner height of 14.6 cm. The cathode is constructed from a wire grid with an approximate height of 4 cm and an approximate width of 3.5 cm (figure 2), and the system features a high voltage (HV) system consisting of a variable autotransformer, neon sign transformer, and Cockcroft-Walton voltage multiplier. In total, the system can produce -33 kV at 10 mA. A rotary vane pump and turbo molecular pump are used to pump down to an ultimate vacuum of 3 mTorr.

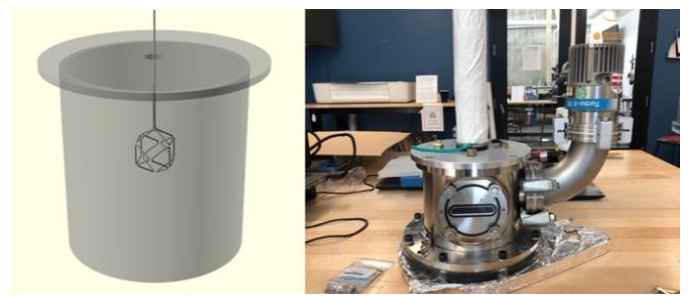

**Figure 2.** 3D model of anode and cathode (left) and vacuum chamber on tabletop (right)

At low pressures using air, the creation of plasma was observed and pressure vs. voltage plots were created from experimental data (figure 3). We attempted to fit the data with the classic Paschen curve, but due to the invariant relative positions of left asymptote and minimum – which are spaced by a factor of e – in the theoretical derivation, we were unable to achieve a good fit.

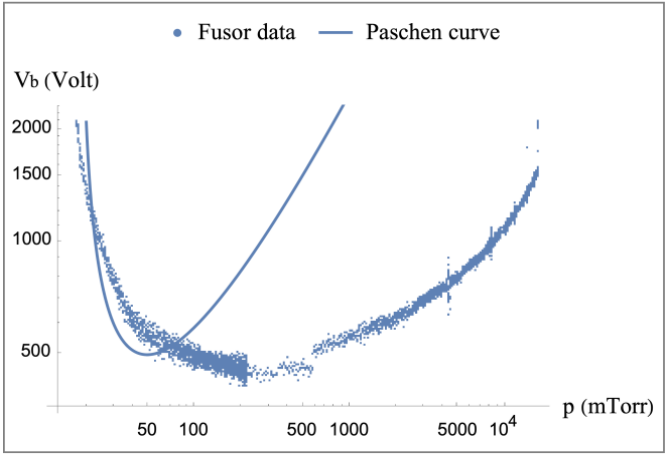

**Figure 3.** Pressure and voltage plot from the Fusor with attempted Paschen curve fit.

The intent of the following Monte Carlo simulations is to allow us to better understand these results. This work contributes to an underexplored area of the plasma physics field, as few computational investigations of electric breakdown between electrodes of geometries other than the parallel plate have been performed. Since many IEC fusion devices more closely align with the concentric sphere and sphere-in-cylinder geometries explored in this paper rather than the parallel plate model, our results are relevant to many fusor engineers and researchers who work with plasma creation process.

**Methods**

In our MC simulations we define new unitless variables $N_i$ and $N_c$, where $N_i = \frac{qV}{U_i}$ and $N_c = \frac{d}{\lambda}$, where $q$ is the charge of an electron, $\lambda$ is the MFP, and $U_i$ is the ionization energy. In other words, $N_i$ is the number of ionization energies an electron could pick up travelling between the cathode and anode, and $N_c$ is number of collision lengths between the anode and cathode. The distance between electrodes, $N_i$, $N_c$, and $U_i$ are used as parameters for the MC while $\lambda$ and V are derived from these quantities. We operate mainly with a fixed distance and $U_i$ value and vary $N_i$ and $N_c$ to search for breakdown. These unitless variables were chosen to simplify our MC simulations by combining multiple quantities into simple variables proportional to V and d.

Upon each simulation, we initialize a stack for tracking electrons to be processed with an initial electron at position 0. For each iteration of the MC, an electron travels a path length, s, which is drawn from the exponential distribution with mean $\lambda$. The new position of the electron is computed, and if there is an ionization (i.e., the electron has a kinetic energy $\geq U_i$), a new electron is pushed to the stack with initial velocity 0 for later processing. After a complete simulation, the mean number of electrons produced per electron during its path from the anode to the cathode is recorded for a starting $N_i$ and $N_c$. To find breakdown values, $N_c$ and $\gamma_e$ are held constant, the average electrons produced per electron are computed over increasing $N_i$, and the voltage at which the function surpasses $1/\gamma_e$ is calculated.

Two languages (Mathematica and Java) were used to cross-verify results. The following subsections will describe techniques used to for each electrode geometry.

*2.1 One-dimensional parallel plate MC*

In this MC, the electron travels solely in the x-direction, so we simply increment the x-coordinate of the electron by the path length to compute the new position.

To improve the time efficiency of the 1D parallel plate MC, weighted method is used. Each electron is initially assigned a weight value of 1. After an ionization, the weight of the electron is doubled – representing the new electron(s) created – and then added to the recorded number of electrons created by that electron. This MC is more computationally efficient as the computation time is not affected by the exponential growth of electrons. For 1D simulations, we calculate alpha by plotting the average number of free electrons created per electron over increasing $N_c$ at a constant $\lambda_i$ value and finding the slope of the resulting curve.

To account for discrepancies between the literature alpha predictions and those computed in our 1D parallel plate MC, we explore a new alpha prediction using a Markov chain. This new prediction models 1D alphas using a state machine. To understand the method, we can use a simplified model with only 6 states, labelled 0-5 (figure 4). Each state describes the energy that an electron has collected, and an electron "moves" from state to state with probability *1-p*, where *p* is the probability of collision. In this example, state 5 represents an electron with ionizing energy, and state 0 represents an electron with 0 energy. The state transitions are modelled as a 6x6 matrix. All electrons start as a vector in state 0, and this vector is multiplied by the transition matrix for each step. This propagates the electron *N* steps through the state machine and then calculates the final count of electrons, $N_e$ (eq. 2).

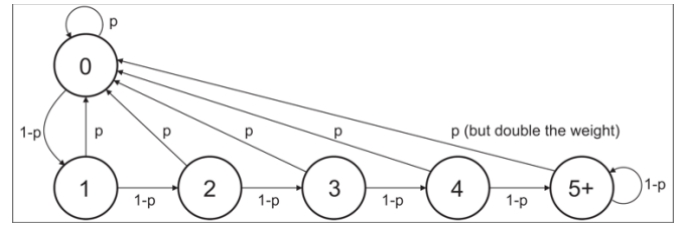

**Figure 4.** MC system as a state machine, where p is the probability of a collision.

$$N_e = \begin{pmatrix} 1 & 1 & 1 & 1 & 1 & 1 \end{pmatrix} \begin{pmatrix} p & p & p & p & p & 2p \\ 1-p & 0 & 0 & 0 & 0 & 0 \\ 0 & 1-p & 0 & 0 & 0 & 0 \\ 0 & 0 & 1-p & 0 & 0 & 0 \\ 0 & 0 & 0 & 1-p & 0 & 0 \\ 0 & 0 & 0 & 0 & 1-p & 1-p \end{pmatrix}^N \begin{pmatrix} 1 \\ 0 \\ 0 \\ 0 \\ 0 \\ 0 \end{pmatrix} \quad (2)$$

As N increases, the eigenvalues of this matrix predict the exponential growth of the ion count. The eigenvalues of the matrix satisfy eq. 3:

$$\lambda^{n+1} - \lambda^n = p(1-p)^n. \qquad (3)$$

Eq. 3 has one solution with $\lambda$ real and greater than 1, which we can approximate by taking the limit as the number of states goes to infinity. From this, we derive eq. 4:

$$\alpha = \frac{1}{\lambda_i} W_0(\lambda_i e^{-\lambda_i}) \qquad (4)$$

Where $W_0$ is the Lambert W function. Using this equation, we derive a new expression for the breakdown voltage (eq. 5):

$$V_B = \frac{Apd + B}{\ln(pd) + C} \qquad (5)$$

This is the same as the Paschen's original model shifted up and to the right.

### 2.2 Three-dimensional parallel plate MC

In the 3D parallel plate MC, to facilitate electron movement and track the coordinates of each electron, we solve for the time an electron takes to travel a determined path length. This is shown in eq. 6, derived from the standard equations of motion along a parabola.

$$s(t) = \int_0^T \sqrt{v_w^2 + (v_x + at)^2} \, dt \qquad (6)$$

Here, $s$ is the pathlength, $a$ is the acceleration in the $x$ direction (from cathode to anode) and $v_w$ is the transverse velocity. The position of the electron is found by updating the coordinates using the calculated time with the electron's tracked velocities and acceleration.

### 2.3 Sphere-in-sphere MC

In this MC, Euler's method is used for tracking the positions and velocities of each electron. The error in the Euler's method calculation is bounded by requiring approximate energy conservation. For this, we reduce $\Delta t$ until $N_c \times rms(error) < \frac{V}{1000}$ and find that further reducing $\Delta t$ does not change our measured results. After an ionization, $U_i$ is subtracted from the original electron's velocity and the original electron scatters. If there is a collision (and no ionization), an energy is drawn from the uniform distribution between $0.1U_i$ and $0.5U_i$ and subtracted from the electron's kinetic energy, representing an approximation of the energy lost as the electron enters a rotational state.

In the concentric-spheres MC, $E(r) = \frac{c_e}{r^2}$, where $c_e = \frac{Vab}{b-a}$ and a and b are the radii of the inner and outer spheres, respectively. This model is derived from the integral in eq. 7, which is evaluated and solved for $c_e$.

$$V = \int_a^b E(r) dr = \int_a^b \frac{c_e}{r^2} dr \qquad (7)$$

### 2.4 Sphere-in-cylinder MC

For the sphere-in-cylinder MC we use the same Euler's method and scattering methods used in sphere-in-sphere. To determine the electric potential between the sphere and the cylinder, the Laplace equation is solved in Mathematica with Dirichlet conditions such that the electric potential equals -1 volts at the anode and 0 volts at the cathode. We can scale the results of the computation to find the electric field at any given point when an arbitrary voltage is applied to the vacuum chamber. We add density to the default mesh in Mathematica to improve the accuracy of the result. We then evaluate this model for the potential at several grid points using a grid spacing of 0.0001 meters (figure 5).

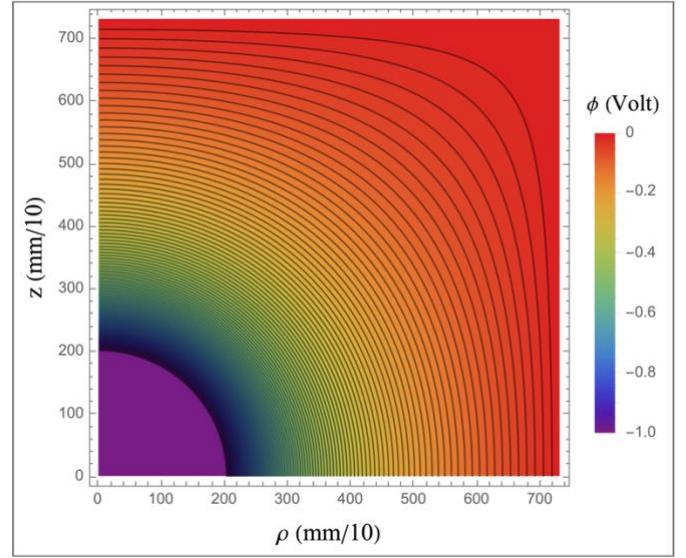

**Figure 5.** Contour plot of the sphere-in-cylinder electric potentials list.

We interpolate between grid points to calculate the electric field in the radial ($\rho$) and vertical ($z$) directions. We chose to use this interpolation system to have consistent potential and E-field values across our Java and Mathematica simulations. For the sphere-in-cylinder geometry with a grid spacing of 0.1 mm, we have found this interpolation system to accurately approximate the direction and magnitude of the field when compared to the continuous electric field function from the Laplace equation. One grid space away from the cathode we observe a maximum of 5% error in the E field magnitude and a maximum of 0.18 degrees of inaccuracy in the E field direction. However, moving away from the cathode, this error decreases significantly to a negligible amount of inaccuracy in magnitude and direction.

In this MC, we use spherical harmonics to approximate the distribution of breakdown voltage over polar angle $\theta$ (figure 6). A uniform distribution over $\theta$ does not approximate the real conditions for plasma creation since electrons emerging from certain angles on the sphere cause breakdown at lower voltages than others. Using spherical harmonics avoids any biases introduced by angular binning of the data. After a given electron is simulated, the number of ionizations and $\theta$ are recorded, addressed as $f_i, \theta_i$, respectively. We do not track $\phi$ and we use we use spherical harmonic $Y_l^m$ values with even values of $l$ from 0 to 10 with $m = 0$ since we assume bipolar and azimuthal symmetry. We approximate the number of electrons created as a smooth function of $\theta$ using spherical harmonics (eq. 8).

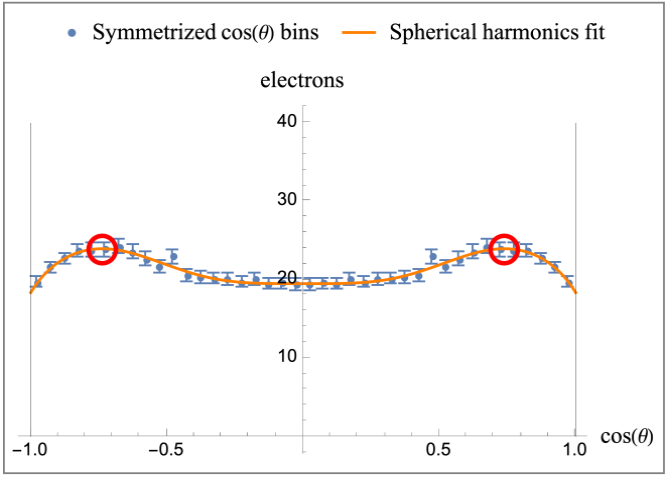

**Figure 6.** Plot of electrons in $\cos(\theta)$ bins with spherical harmonics fit at $\lambda^{-1} = 1 \text{ cm}^{-1}$ and $N_i = 10$. The red circles represent the $\cos(\theta)$, $\pm 0.738$, that yielded the max electron count.

$$f(\theta) \simeq \sum_{l=0}^{5} c_{2l} Y_{2l}^0(\theta, 0) \quad (8)$$

## Results

### 3.1 One-dimensional parallel plate MC

Figure 7 demonstrates the tendency for the literature prediction of alpha to overestimate alpha at lower $\lambda_i$ values, compared to our results. The Markov chain prediction agrees with the MC results for 1D alphas.

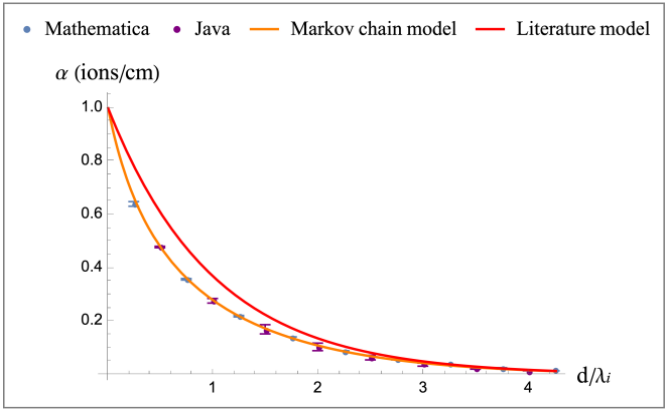

**Figure 7.** 1D MC alpha values and models compared

Figure 8 demonstrates the skewing of the sample space which affects literature alpha predictions, shown in the following step from the alpha derivation (eq. 9).

$$\frac{\exp\left(-\frac{\lambda_i}{\lambda}\right) - \exp\left(-\frac{x}{\lambda}\right)}{1 - \exp\left(-\frac{x}{\lambda}\right)} \approx \exp\left(-\frac{\lambda_i}{\lambda}\right) \quad (9)$$

The literature derivation of alpha shown in eq. 9 assumes that x, the interval in which the growth of the electron count is analyzed, is significantly larger than $\lambda$. Alpha values from the standard alpha prediction assume that the electron creation rate is invariant with x, but in a more realistic model the sample space of x is skewed by exponential growth of electrons (figure 9). Especially at lower pressures and higher voltages, this skews the electrons that have arrived at x to be less likely to have travelled $\lambda_i$, meaning there are fewer ionizations in a more realistic model than in the literature derivation.

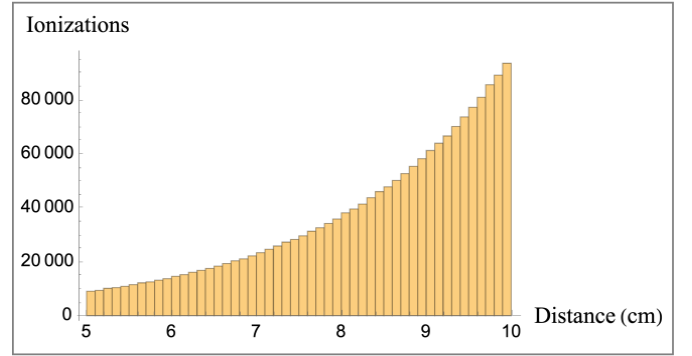

**Figure 8.** Observation of sample space for alpha calculation. 10,000 reps performed at x = 5, $N_c$ = 10, and $N_i$ = 20.

Figure 9 shows that the 1D breakdown curve has a lowered minimum compared to the literature curve. This discrepancy can be attributed to the overestimation of alpha by the literature model at select $\lambda_i$ values which correspond to the Nc values around the minimum of the curve. Due to the skewing of sample space in our simulations, our alphas are consistently smaller in this range, meaning the system needs more voltage to produce the number of ions required for ionization. Despite this disagreement, the MC results and the literature model agree at the left- and right-hand asymptotes. In addition, the Markov chain-derived model appears to improve upon the original model in fitting the MC results, especially near the minimum. However, more works needs to be done to accurately approximate the minimum of the curve since there is still a slight discrepancy.

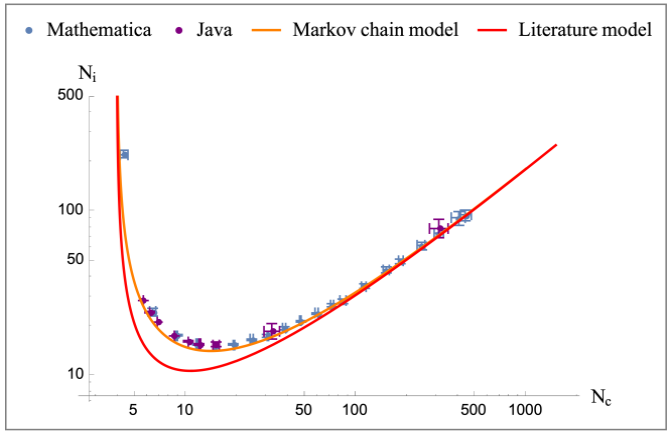

**Figure 9.** 1D parallel plate breakdown for $\gamma_e = 50$.

### 3.2 Three-dimensional parallel plate MC

Figure 10 shows that the 3D breakdown MC data has a slightly lowered minimum compared to the literature curve, although it fits more closely with Paschen's original model than the 1D MC data.

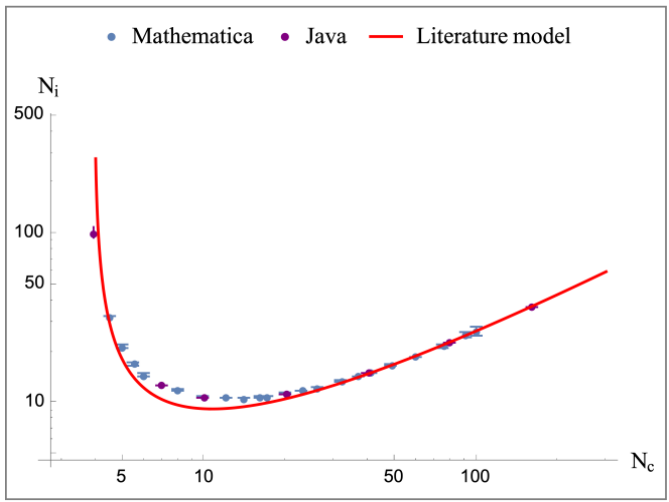

**Figure 10.** 3D parallel plate breakdown values for $\gamma_e = 50$.

### 3.3 Sphere-in-sphere MC

Figure 11 confirms that the concentric-spheres breakdown curve features a broadened shape with slower growth at larger Nc with a slightly shifted minimum compared to the parallel-plate curve. There is some agreement at the left-hand asymptote.

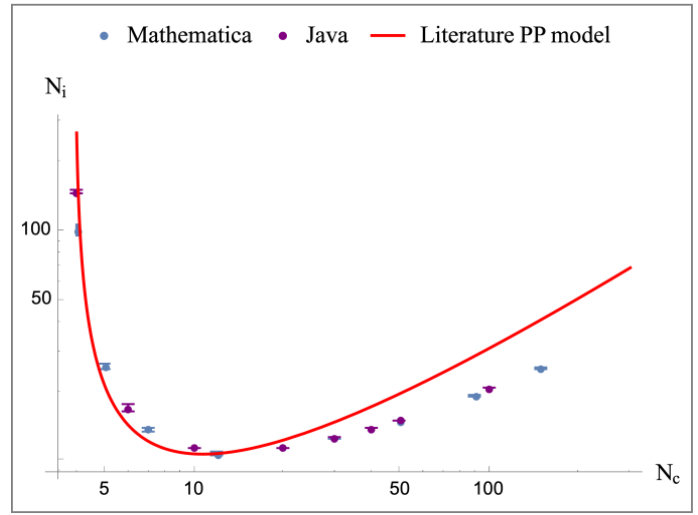

**Figure 11.** Concentric spheres breakdown for $\gamma_e = 50$.

### 3.4 Sphere-in-cylinder MC

Figure 12 shows that the sphere-in-cylinder MC breakdown results agree with the sphere-in-sphere breakdown results at higher pd values, but also feature a different shape and behavior at the left-hand asymptote, where the curve grows more slowly compared to the sphere-in-sphere curve.

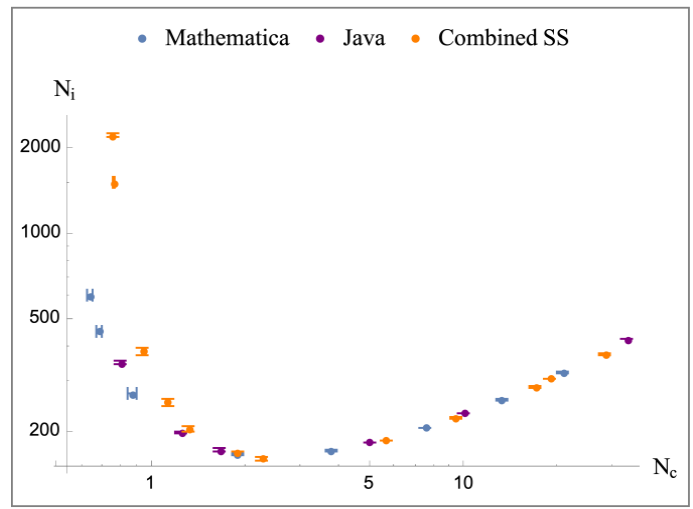

**Figure 12.** Sphere-in-cylinder breakdown for $\gamma_e = 50$. "Combined SS" is the sphere-in-sphere breakdown results from Java and Mathematica combined.

Figure 13 helps to explain the behavior at the left asymptote: at lower Nc values, the electrons launched on the diagonals towards the corners had more chances to collide with ions and kickstart the Townsend avalanche due to a longer distance of travel. The difference in ionizations across the launching angles is more noticeable at lower Nc values than at higher Nc values, which helps to explain why it agrees with the concentric spheres curve at higher Nc.

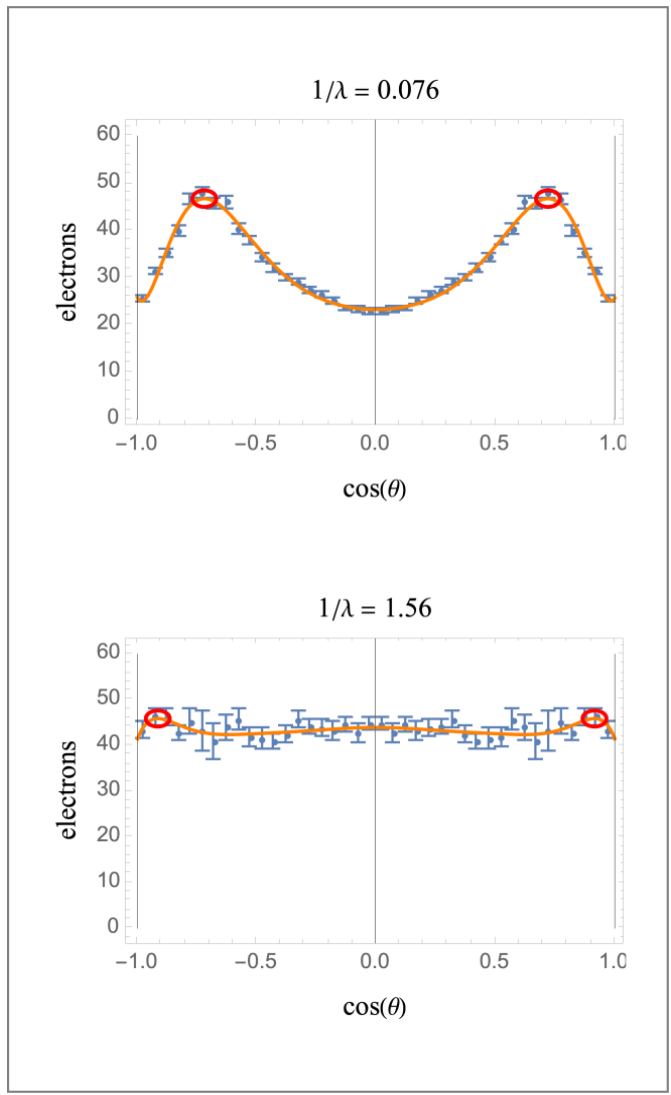

**Figure 13.** Plot of electrons in cos(θ) bins with spherical harmonics fit at two different pressures in the sphere-in-cylinder case.

## Discussion

We had expected to observe a widened curve for concentric spheres and sphere-in-cylinder geometries due to the widened nature of the curve for our experimental setup, which more closely resembled those geometries as opposed to parallel plates. In this paper we have verified this prediction and found surprising behavior of the sphere-in-cylinder curve at the left-hand asymptote. We were surprised by the discrepancies between our parallel plate results and those of the literature especially at lower $\lambda_i$ values where our simulations consistently calculate less exponential growth of electrons, raising the minima and part of the left-hand asymptotes for the MC-derived curves. We are optimistic about the Markov chain-derived model and its improvements in predicting our MC breakdown results, but more work must be done to understand the shortcomings of the model.

Our largest source of error is the simplification of 3D physics in the simulations. More specifically, the electrostatic repulsion and attraction between ions and electrons in the systems is not accounted for, and the collision cross-section between electrons and gas molecules was assumed to be a constant over varying energies.

By exploring more uncommon electrode geometries such as the concentric spheres geometry and sphere-in-cylinder geometry, these findings fill a key gap in the scientific literature. However, our findings follow the pattern we observed in other recent notable publications addressing Paschen's law: Paschen's original model is insufficient to fit our data, but a modified curve following the general shape is applicable. Further work might continue to explore the Markov chain model and try to improve the fit at the minimum of the parallel plate breakdown curve. Other geometries, gasses, and temperatures may also be explored. Results for these simulations may be compared to experimental data to explore the accuracy and applicability of MC results.

## Conclusion

Our simulations support our claim that alpha values as predicted by the literature model will be overestimated due to the skewing of the sample space from the exponential growth of ionizations, which is exacerbated at large voltages relative to pd. This mainly affects the minimum of the 1D breakdown curve, and we have found that a Markov chain derived model for breakdown more closely matches MC results at the minimum. We confirm that the concentric sphere and sphere-in-cylinder geometries feature a broadened breakdown curve, which matches results from our experimental fusor setup.

## Acknowledgements

We thank our advisors Dr. Charles Whitmer and Gunnar Mein for their contributions to our research and for their support of the fusion reactor project at Eastside Preparatory School over the past seven years.